\documentclass[conference]{IEEEtran}
\usepackage[T1]{fontenc}
\IEEEoverridecommandlockouts
\usepackage{cite}
\usepackage{amsmath,amssymb,amsfonts}
\usepackage{algorithmic}
\usepackage{graphicx}
\usepackage{textcomp}
\usepackage{xcolor}
\usepackage[hyphens]{url}
\usepackage[hidelinks]{hyperref}
\hypersetup{breaklinks=true}

\def\BibTeX{{\rm B\kern-.05em{\sc i\kern-.025em b}\kern-.08em
    T\kern-.1667em\lower.7ex\hbox{E}\kern-.125emX}}
\begin{document}

\title{Datacenters in the Desert: Feasibility and Sustainability of LLM Inference in the Middle East
}
% \author{
%     \IEEEauthorblockN{Anonymous Authors}
%     \IEEEauthorblockA{
%         % \textit{Paper under double-blind review}
%     }
% }
\author{
\IEEEauthorblockN{Lara Hassan\textsuperscript{*}}
\IEEEauthorblockA{\textit{Department of Computer Science} \\
\textit{MBZUAI} \\
Abu Dhabi, UAE \\
lara.hassan@mbzuai.ac.ae}
\and
\IEEEauthorblockN{Mohamed ElZeftawy\textsuperscript{*}}
\IEEEauthorblockA{\textit{Department of Computer Science} \\
\textit{MBZUAI} \\
Abu Dhabi, UAE \\
mohamed.elzeftawy@mbzuai.ac.ae}
\and
\IEEEauthorblockN{Abdulrahman Mahmoud}
\IEEEauthorblockA{\textit{Department of Computer Science} \\
\textit{MBZUAI} \\
Abu Dhabi, UAE \\
abdulrahman.mahmoud@mbzuai.ac.ae}
\thanks{\textsuperscript{*}Equal contribution.}
}

\maketitle

\begin{abstract}
As the Middle East emerges as a strategic hub for artificial intelligence (AI) infrastructure, the feasibility of deploying sustainable datacenters in desert environments has become a topic of growing relevance. This paper presents an empirical study analyzing the energy consumption and carbon footprint of large language model (LLM) inference across four countries: the United Arab Emirates, Iceland, Germany, and the United States of America using DeepSeek Coder 1.3B and the HumanEval dataset on the task of code generation. We use the CodeCarbon library to track energy and carbon emissions and compare geographical trade-offs for climate-aware AI deployment. Our findings highlight both the challenges and potential of datacenters in desert regions and provide a balanced outlook on their role in global AI expansion.

\end{abstract}

\begin{IEEEkeywords}
Sustainable AI, Datacenter Efficiency, Carbon Emissions, LLM Inference
\end{IEEEkeywords}
\section{Introduction}
With the explosion of large-scale artificial intelligence workloads, the environmental footprint of datacenters has come under scrutiny. The AI compute coming online appears to be increasing by a factor of 10 every six months. This challenge is particularly acute in regions such as the Middle East, where extreme temperatures and fossil-based grids pose energy efficiency challenges. The United Arab Emirates (UAE) and Kingdom of Saudi Arabia (KSA) have announced multi-gigawatt AI datacenter projects \cite{pwc2025datacentre}, backed by renewable energy investments and global partnerships. This paper investigates whether these desert-based infrastructures can compete in sustainability with traditional cold-climate hubs.
\section{Background}

Recent developments highlight the Middle East's accelerating role in global AI infrastructure. A landmark report by SemiAnalysis~\cite{semianalysis2025ai} outlines major trilateral agreements between USA, UAE, and KSA to deploy cutting-edge datacenter capacity across the region. One such deal involves Abu Dhabi-based G42 receiving an annual quota of 500{,}000 NVIDIA GPUs, with 20\% retained locally, as part of a plan to develop a 5~GW AI campus powered by a mix of solar, gas, and nuclear energy.

These investments are also economically strategic. Power costs in the Middle East remain significantly lower than in Western markets, with some solar plants in Abu Dhabi producing electricity at prices as low as \$0.014 per kWh. This affordability enables cost-effective operation of large-scale AI workloads. However, the region still faces considerable environmental challenges, particularly the high cooling demands imposed by extreme temperatures. Addressing these challenges has become a focal point of regional infrastructure planning and innovation efforts.

One of the key metrics for datacenter efficiency is the Power Usage Effectiveness (PUE), which measures the ratio of total facility energy to IT equipment energy. A lower PUE indicates higher efficiency, as it means a greater proportion of the energy is being used directly for computing rather than for cooling, lighting, or other overhead. In desert climates, extreme ambient temperatures significantly burden traditional air-cooled systems, often pushing PUE values above 1.8 due to increased cooling demands. Alternative cooling methods such as evaporative, liquid immersion, or seawater-based systems offer improved thermal efficiency, with PUE values approaching 1.3--1.5, but introduce trade-offs in water usage and system complexity. The Middle East is actively investing in advanced cooling technologies to improve operational efficiency under harsh climatic conditions \cite{mordor2025middle}. To further enhance sustainability, new datacenter initiatives are incorporating on-site solar power generation. These developments reflect a broader shift toward environmentally responsible AI infrastructure in arid regions.

\section{Methodology}
We conduct inference using the DeepSeek Coder 1.3B model~\cite{guo2024deepseekcoderlargelanguagemodel} on the HumanEval dataset~\cite{chen2021evaluatinglargelanguagemodels}, targeting the code generation task. All experiments are performed on a consistent hardware setup that includes an NVIDIA RTX 5000 ADA Generation GPU and an Intel(R) Xeon(R) w7-2495X CPU. To evaluate the environmental impact of LLM inference across different geographies, we use the open-source CodeCarbon library \cite{benoit_courty_2024_11171501}, which estimates energy consumption (in kWh) by monitoring CPU, GPU and RAM power usage and calculates carbon emissions (in kgCO$_2$) based on the carbon intensity of the regional energy grid. These estimations rely on publicly available datasets valid up to the year 2023 that reflect the energy mix and emission factors of each location. We simulate inference in four distinct regions: UAE, Iceland, Germany, and Texas, USA, chosen for their diversity in energy infrastructure and climate. The UAE represents a desert climate with a grid dominated by natural gas and emerging solar and nuclear inputs; Iceland reflects a subpolar climate powered almost entirely by renewable geothermal and hydroelectric sources; Germany features a temperate climate with a mixed energy grid combining renewables and fossil fuels; and Texas embodies a hot, semi-arid to humid subtropical climate with a diverse energy mix that includes significant wind and natural gas contributions, as well as growing solar capacity. By holding the workload and hardware constant across all simulations, we isolate the effect of geographic factors on carbon metrics. We report total energy consumed and associated CO$_2$ emissions, which are directly influenced by the carbon intensity of each region's energy grid.

\section{Results}
\begin{figure}[h!]
    \centering
    \includegraphics[width=1\linewidth]{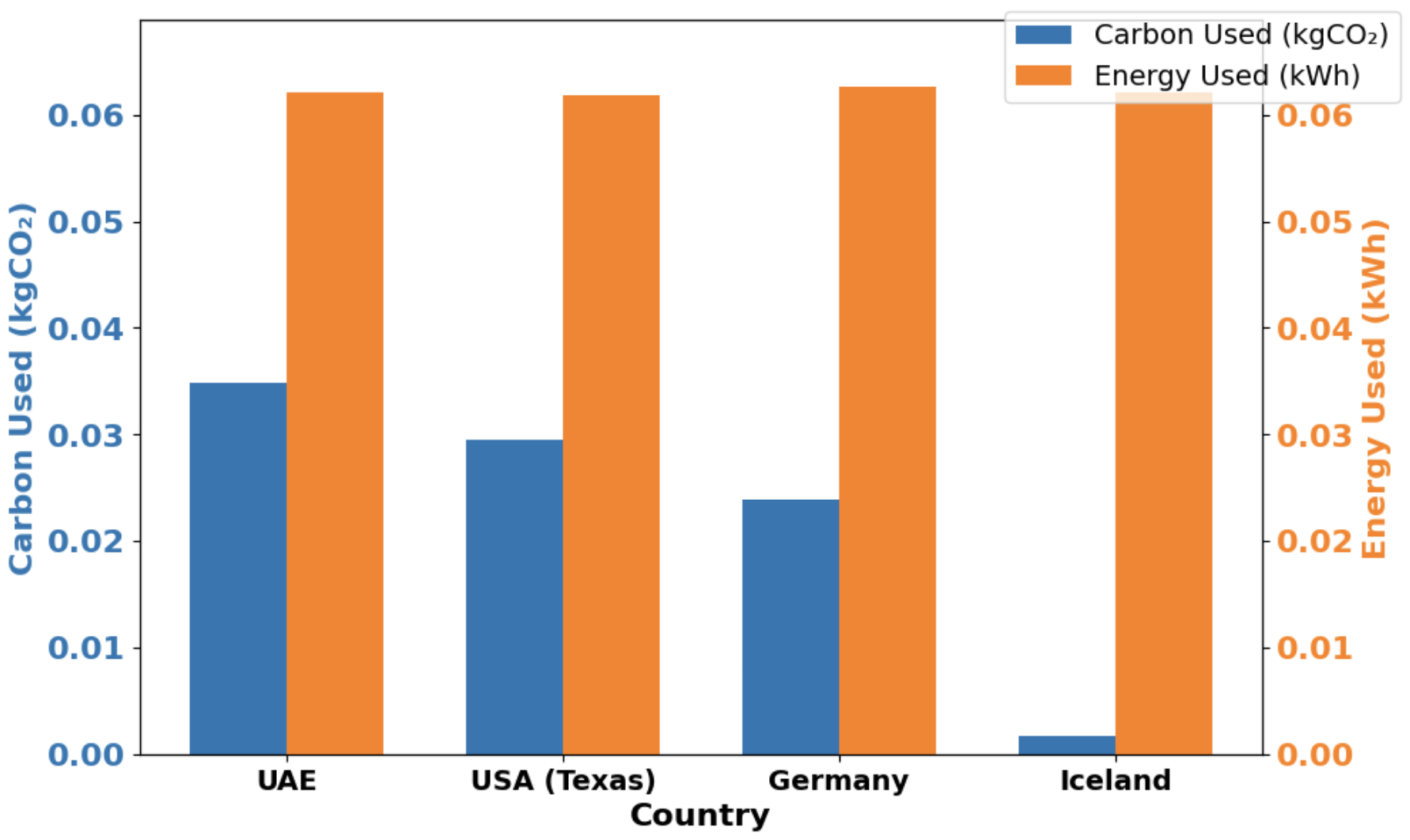}
    \caption{Carbon and energy usage for equivalent inference workloads across regions.}
    \label{fig:carbon_energy}
\end{figure}

Figure~\ref{fig:carbon_energy} presents a comparative analysis of carbon emissions (in kgCO$_2$) and total energy usage (in kWh) for equivalent LLM inference workloads across four regions: the UAE, Iceland, Germany, and Texas, USA.

\textbf{Carbon Emissions.} As shown in Figure \ref{fig:carbon_energy}, differences in carbon emissions were significant. Although energy usage was constant through the experiments, the associated CO$_2$ emissions in the UAE and Texas were drastically greater due to the fossil fuel-heavy energy mix. The UAE’s simulated inference emitted slightly more CO$_2$ than Texas and orders of magnitude more than Iceland, which has a near 100\% renewable energy mix. Germany, with its partially decarbonized grid, produces moderately lower emissions than Texas and the UAE. These results underscore the critical importance of energy source as emissions are driven far more by grid composition than by efficiency alone.

\textbf{Cost of Electricity.} When considering operational costs, the UAE emerges as a strong candidate due to its significantly lower electricity price. At just \$0.077 per kWh, the UAE offers the lowest cost among the four regions evaluated compared to \$0.109 in the USA (Texas), \$0.156 in Iceland, and \$0.323 in Germany. This means that despite higher carbon emissions, the UAE achieves notably lower cost per inference. For organizations scaling LLM inference at massive volumes, this economic advantage can be a decisive factor, potentially outweighing the environmental drawbacks when emissions can be offset through clean energy procurement.

\section{Discussion}
%% OLD SECTION TO BE ADDED AGAIN IF THE NEW TRIAL IS MESSED UP

% Our analysis reveals that while datacenters in the UAE face higher carbon emissions due to the region's climate and energy mix, the significantly low electricity cost (\$0.077/kWh) offers a major economic advantage for large-scale LLM inference, highlighting a trade-off between sustainability and cost efficiency. 
% Our analysis reveals that while datacenters in the UAE and Texas incur higher carbon emissions due to their regional energy mixes and climatic conditions, their significantly lower electricity costs present a compelling economic advantage for large-scale LLM inference. This underscores a critical trade-off between environmental sustainability and cost efficiency.

Our findings indicate that datacenter deployments in regions like the UAE and Texas result in higher carbon emissions compared to locations such as Iceland and Germany, primarily due to fossil-fuel-dominated energy grids. However, these regions simultaneously offer substantial economic advantages for large-scale LLM inference due to significantly lower electricity prices, highlighting a key trade-off between cost-efficiency and environmental sustainability.

From a global infrastructure perspective, there are compelling reasons to view the Middle East's datacenter future with cautious optimism. Countries in the Middle East, such as the UAE and KSA, are aggressively investing in clean energy infrastructure. Co-locating datacenters with solar and nuclear power facilities can drastically reduce the carbon intensity of operations. Solar energy aligns well with cooling demand, and nuclear provides reliable baseload power, creating a favorable long-term sustainability trajectory.

In parallel, the region is adopting advanced cooling technologies and infrastructure best practices to enhance energy efficiency. Hot/cold aisle containment, liquid cooling, and AI-optimized HVAC systems have shown measurable improvements, with PUE reductions exceeding 0.4 in local deployments. Major cloud and colocation providers in the Middle East are now targeting PUE values below 1.5, aligning with international benchmarks. 

That said, similar progress is being made in regions like Texas, where renewable energy penetration (particularly wind and solar) continues to grow with competitive electricity pricing and an established digital infrastructure ecosystem. This makes it an attractive location for scalable and cost-efficient AI compute, especially for organizations prioritizing both economic performance and gradual decarbonization.

% Furthermore, the global demand for AI compute is rapidly growing, while many Western regions face infrastructure saturation and regulatory complications. In contrast, Middle Eastern countries' ability to build capacity quickly and to do so cost-effectively positions them as valuable players in the global AI infrastructure ecosystem. By 2030, the Middle East is projected to contribute over 6~GW of datacenter capacity. If coupled with continued investment in clean energy, these desert facilities can combine economic viability with acceptable sustainability performance.
The global demand for AI compute continues to grow rapidly, placing increasing pressure on existing infrastructure across multiple regions. In some Western countries, expansion can be constrained by regulatory processes, land availability, and grid limitations. Emerging markets, including parts of the Middle East, have demonstrated the ability to deploy datacenter capacity at a relatively fast pace and at competitive costs. Projections suggest the region may contribute over 6~GW of additional capacity by 2030. When paired with ongoing investments in clean energy and advanced cooling solutions, datacenters in desert environments could offer a complementary option within a diversified global AI infrastructure strategy balancing economic efficiency with evolving sustainability goals. Moreover, geographical diversification in datacenter placement increases global resilience, improves latency to fast-growing markets, and supports a more distributed AI infrastructure backbone.

% These factors make the Middle East a strategic and increasingly sustainable location for future datacenter investments. 
Looking ahead, in the next years we are likely to see increased integration of on-site renewable energy, particularly solar, coupled with advancements in modular and liquid-based cooling systems. Policy incentives and regional cooperation may further accelerate low-carbon datacenter initiatives. However, trade-offs in water consumption, infrastructure cost, and land use efficiency will require careful evaluation to ensure that the region’s datacenter growth aligns with broader environmental and economic goals.

\section{Conclusion}

Our findings highlight a fundamental trade-off in the global deployment of LLM inference workloads: while regions like Iceland offer unmatched sustainability due to their clean energy mix and natural cooling advantages, they come at a higher cost per kilowatt-hour. In contrast, UAE and USA present an economically attractive alternative, with the lowest electricity cost among the studied regions. However, this affordability is currently accompanied by higher carbon emissions, primarily due to the fossil-fuel-dominated grid.

Sustainable LLM deployment in the Middle East is already underway, driven by substantial investments in solar and nuclear energy, coupled with the adoption of advanced cooling technologies. The question now is not whether it can happen, but how to ensure it is done responsibly by maximizing energy and carbon efficiency while navigating trade-offs in water use, infrastructure complexity, and environmental impact. With strategic planning, supportive policies, and continued innovation, datacenters in desert environments can achieve a viable balance between economic competitiveness and sustainability.

\bibliographystyle{IEEEtran}
\bibliography{references}

\end{document}